\documentclass[%
 reprint,
 amsmath,amssymb,
 aps,
]{revtex4-2}

\usepackage{graphicx}% Include figure files
\usepackage{dcolumn}% Align table columns on decimal point
\usepackage{amsmath}
\usepackage{bm}% bold math
\usepackage{float}
\begin{document}

\preprint{APS/123-QED}

\title{Experimental Evidence for Longitudinal Scaling Exponent Saturation in Shear Turbulence }%

\author{Dipendra Gupta}
 \email{dg535@cornell.edu}
 
\author{Gregory P. Bewley}%
\affiliation{%
 Sibley School of Mechanical and Aerospace Engineering, Cornell University, Ithaca, NY 14850, USA\\}

\date{\today}% It is always \today, today,

\begin{abstract}

The asymptotic behavior of velocity statistics in the tails of distributions and at high Reynolds numbers remains unresolved in turbulence. 
To investigate this behavior we measured the $n$th-order moments of the distributions of longitudinal velocity differences, 
$S_n(r) \equiv \langle [u(x+r)-u(x)]^n \rangle \sim r^{\zeta_n}$, 
in turbulent shear layers at Taylor-scale Reynolds numbers up to $Re_\lambda \approx 1400$. 
We used a nanoscale hot-wire probe with a sensing length, $l_w$, that was about half the Kolmogorov scale, $\eta$. % ($l_w/\eta \approx 0.5$). 
We obtained datasets that were up to $5\times 10^7$ integral timescales long, 
so that the statistics converged up to $n=14$. 
In the inertial range, the exponents, $\zeta_n$, deviate from classical models and appear to saturate near $\zeta_n \approx 2.2 \pm 0.1$ for $n \gtrsim 12$. 
The saturation in the exponents is supported by a collapse of the tails of the velocity-difference distributions, and by plateaus in their compensated moments. 
These results constitute the first experimental evidence for scaling exponent saturation in longitudinal velocity increments, and is consistent with a dominance of localized vortex filaments in turbulence. 

\end{abstract}

\maketitle

%\section{Introduction}

Turbulent flows exhibit a hierarchy of interacting scales, from large energy-containing motions to small dissipative structures. 
This multiscale dynamics is a feature of the Navier–Stokes equations, 
whose nonlinearities mediate kinetic energy exchange between different scales.  
A probe of motion on different scales is the velocity increment, 
$\Delta u \equiv u(x+r, t) - u(x, t)$, and its statistics, 
including the moments, or structure functions, 
$S_n(r) \equiv \langle (\Delta u)^n \rangle$, 
where the brackets, $\langle\cdot\rangle$, denote a time average in our experiments. 
For intermediate scales, $r$, whose dynamics are dominated by inertia, 
Kolmogorov’s 1941 theory predicts $S_n(r) \sim r^{\zeta_n}$ with $\zeta_n = n/3$ for homogeneous and isotropic turbulence \cite{kolmogorov1941local}. 
Observed deviations from this scaling reflect the increasingly intermittent distributions of large velocity differences toward smaller scales, and are now well established \cite{frisch1995turbulence,ishihara2009study,dubrulle2019beyond}. 
The intermittency reflects the propensity of turbulence to generate extreme events at small scales, and increasingly so at higher Reynolds numbers \cite{sreenivasan1997phenomenology,biferale2003shell,johnson2017turbulence,eyink2025beyond}. 

Within the multifractal framework, intermittent turbulent fluctuations are composed of a continuum of singular sets, each characterized by a local Hölder exponent $h$ and a fractal dimension $D(h)$ describing the subset of the flow with that exponent~\cite{frisch1995turbulence, frisch1985singularity}, 
yielding $\zeta_n = \underset{h}{\min}~[nh + 3 - D(h)]$. 
The large $n$ behavior of $\zeta_n$ is controlled by the most singular structures in the flow, which are associated with $h_{\min} \equiv \min(h)$. 
If $h_{\min} > 0$, the exponents scale with $n$ asymptotically as $\zeta_n \sim n h_{\min}$ and grow without bound. 
On the other hand, if $h_{\min} = 0$, $\zeta_n$ saturates to a finite value, $\zeta_\infty = 3 - D(0)$. 
If saturation does occur, it reflects the dominance of spatially localized coherent structures, with the value of the exponent being related to the geometry of the structure. 
For instance if D(0) = 1, then $\zeta_\infty = 2$ and the structures are vortex filaments \cite{frisch1995turbulence,buaria2023saturation}. 
Visualizations from laboratory experiments and numerical simulations suggest that the structures are indeed intense vortex filaments~\cite{douady1991direct,she1990intermittent,jimenez1993structure}. 
Establishing whether longitudinal structure function exponents saturate toward high orders is therefore a fundamental problem, as it constrains the geometry, universality, and dynamical role of the most singular structures in turbulent flows. 

For structure functions of passive scalars in turbulence, such as for ink in water, 
the saturation of the exponents at high orders is well established and is commonly attributed to the presence of sharp, cliff-like features in the spatial distribution of the scalar~\cite{celani2000universality,moisy2001passive,iyer2018steep,warhaft2000passive}. 
A similar phenomenon arises in Burgers turbulence, or in turbulence with no pressure force, where the scaling exponents for velocity differences saturate due to the formation of shocks~\cite{bec2007burgers}. 
For real turbulence, however, the issue of whether exponents saturate remains unresolved. 
Experimental measurements of transverse velocity increments 
have reported saturation at $\zeta_\infty \approx 2.8$ for $Re_\lambda \approx 650$~\cite{staicu2003small}, 
while high-resolution direct numerical simulations report $\zeta_\infty \approx 2$ at $Re_\lambda \approx 1300$~\cite{buaria2023saturation}. 
These results are restricted to transverse statistics, and no evidence of saturation has been established for longitudinal structure functions. 
A recent theory predicts that longitudinal exponents may saturate at $\zeta_\infty \approx 7.3$ for $n \to \infty$~\cite{sreenivasan2021dynamics}, implying a fractal dimension $D(0)=-4.3$, a negative value indicating that the associated singularities are rarer than isolated points and exist only in an ensemble sense \cite{mandelbrot1990negative,sreenivasan1991fractals}. However, this asymptotic regime, which requires arbitrarily high-order moments and probes regions of negative fractal dimension, remains beyond current experimental and numerical accessibility.

Progress on longitudinal velocity statistics at high orders is constrained by the concurrent requirements of large $Re_\lambda$ in order to establish an inertial range, long datasets for statistical convergence at high orders, and sub-Kolmogorov spatial resolution to avoid probe-induced attenuation of extreme events. 
No prior experiment has satisfied all three conditions simultaneously. 
In our experiments, we observed shear turbulence instead of homogeneous and isotropic turbulence in part to reach larger Reynolds numbers than would otherwise be possible in the same facility, and in part 
because it appears that sustained mean gradients amplify strong fluctuations and enhance small-scale intermittency \cite{toschi1999intermittency,gualtieri2002scaling}. 
The continuous deformation of the flow by the mean shear promotes the development of coherent structures \cite{brown1974density,smits2011high,dong2017coherent} and increases the intensity of turbulent fluctuations relative to many other means of generating turbulence in the laboratory \cite{shen2002longitudinal,pumir1996turbulence,sekimoto2016direct}, which extends the experimentally accessible Reynolds numbers to larger values. 
This behavior has the potential of making asymptotic intermittent behavior visible at lower statistical orders than in homogeneous and isotropic turbulence, thereby requiring shorter datasets than would be required for higher order moments to converge statistically. 

In this Letter, we report measurements of longitudinal velocity increments in turbulent shear layers at high enough Reynolds numbers to identify scaling in the inertial range, with datasets long enough for statistics to converge up to the 14-th order, and with a probe small enough to resolve scales smaller than the Kolmogorov scale. 
The scaling exponents saturate to a value ($\zeta_n \approx 2.2$) 
that is not compatible with existing theoretical predictions \cite{sreenivasan2021dynamics}. 
To our knowledge, this constitutes the first experimental evidence for the saturation of longitudinal structure-function exponents in three-dimensional turbulence, and it constrains the geometry of the most singular structures governing the velocity field.

%\section*{Experiments and flow details}

We conducted the experiments in the Warhaft Wind and Turbulence Tunnel at Cornell University (test section $0.91 \times 0.91 \times 9.1$\,m$^3$)~\cite{yoon1990evolution}. 
A statistically stationary planar shear layer was generated by imposing a velocity difference of 6\,m/s between the upper and lower halves of a passive grid, which we achieved by covering the upper half with a fine wire mesh. 
Downstream of the grid, a 0.6\,m long flow straightener composed of horizontal sheets spaced by 0.025\,m drives the mean flow to be approximately parallel. 
Downstream of the flow straightener, the upper and lower flows were separated by a 1.2\,m long splitter plate. 
Measurements were acquired 5\,m downstream of the end of the splitter plate \cite{gupta2023experimental}. 
Temporal signals were converted to spatial statistics using Taylor’s frozen-turbulence assumption, $r = \langle U \rangle \tau$, with $u(t)=U(t)-\langle U \rangle$ denoting velocity fluctuations. 
The turbulence intensity was $\sqrt{\langle{u^2}\rangle}/\langle{U}\rangle \approx 0.15$, within commonly accepted bounds ($\sim 0.1$–$0.2$)~\cite{lumley1965interpretation,tennekes1972first,shen2000anisotropy}, although some bias in high-order statistics cannot be excluded. 
Prior studies indicate that inertial-range exponents are affected at the level of $\lesssim 5\%$ for $\sqrt{\langle{u^2}\rangle}/\langle{U}\rangle \lesssim 0.3$ and moment orders up to $n \lesssim 14$~\cite{pinton1994correction,anselmet1984high}, suggesting that the present conclusions remain robust. 
The Kolmogorov scale was estimated as $\eta = (\nu^3/\epsilon)^{1/4}$, with $\epsilon = 15\nu \langle (\partial u/\partial x)^2 \rangle$, yielding $L/\eta \approx 2500$, where $L$ is the integral scale. 
The corresponding Taylor-scale Reynolds number is $Re_\lambda = \sqrt{\langle u^2 \rangle } \lambda/\nu \approx 1400$, with $\lambda = \left(\langle u^2 \rangle / \langle (\partial u/\partial x)^2 \rangle \right)^{1/2}$.

We measured the velocity with a nanoscale thermal anemometry probe ($l_w \approx 60~\mu\mathrm{m} \approx 0.5\eta$, thickness $0.1~\mu\mathrm{m}$) fabricated at the Cornell NanoScale Facility \cite{gupta2026experimental,liu2021nanoscale}, operated with a constant-temperature anemometer. 
The sub-Kolmogorov resolution reduces spatial filtering and enables the capture of extreme velocity increments \cite{bailey2010turbulence,hutchins2009hot}. 
Signals were acquired continuously over 10 days at $40~\mathrm{kHz}$ using a 16-bit system, yielding $\mathcal{O}(10^7)$ integral time scales. 
Temperature drift was monitored and corrected \cite{smits1983constant}, and signals were analog- and digitally-filtered to suppress noise and aliasing. 
A conventional hot wire was also used for cross-validation at a lower Reynolds number. 
This combination of high Reynolds number, sub-Kolmogorov resolution, and exceptional large dataset overcomes the longstanding experimental constraints on high-order structure functions, enabling converged estimates into the extreme-event regime.

%\subsection{Small-scale statistics of velocity increments measured with the nanoscale hot-wire at $Re_\lambda \approx 1400$}
\begin{figure} %[H]
 \centering
 \includegraphics[width=1.1\linewidth, trim={50 92 50 125},clip]{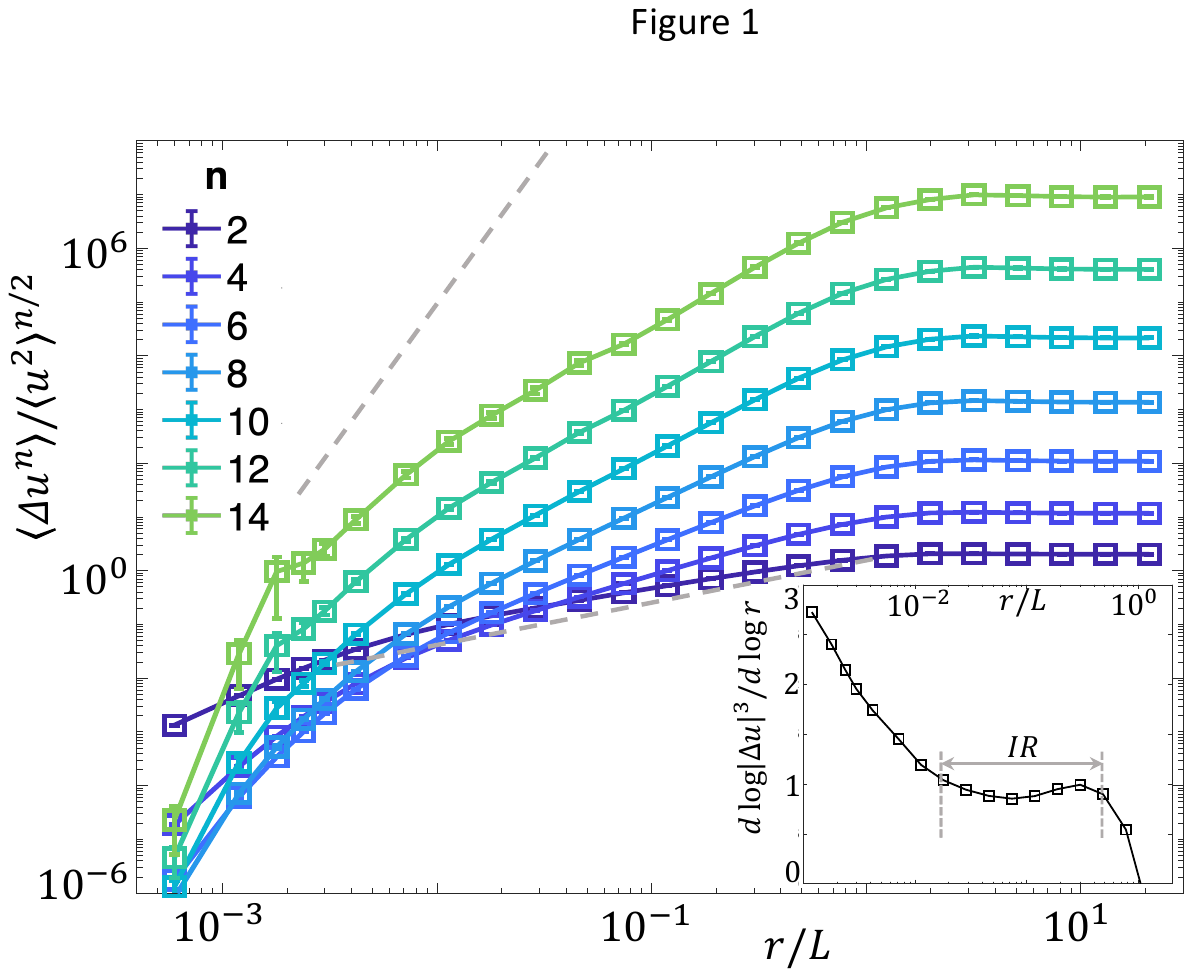}
    \caption{Structure functions $\langle\Delta u^n\rangle$ are normalized by $\langle u^2\rangle^{n/2}$ at $Re_\lambda \approx 1400$. 
The dashed grey lines are K41 predictions \cite{kolmogorov1941local} for $n=2$ (lower) and $n=14$ (upper). 
The inset shows the local slope of the third order structure function, $\zeta_3(r)$ as a function of scale separation. 
IR denotes the inertial range over which an approximate plateau is observed. } 
   \label{fig:Figure1}
\end{figure}

\begin{figure}%[H]
 \centering
 \includegraphics[width=1\linewidth, trim={18 170 15 152},clip]{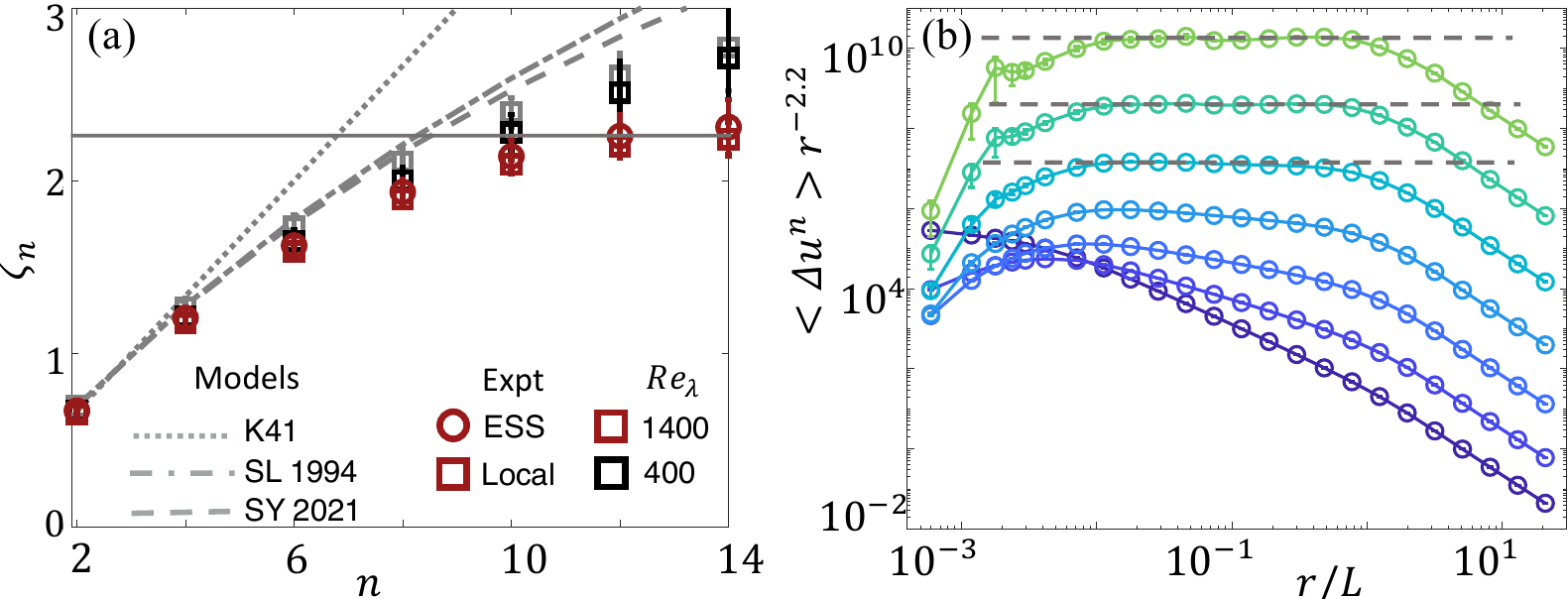}
    \caption{(a) Scaling exponents of even-order longitudinal structure functions at high $Re_\lambda \approx 1400$ (red) and low $Re_\lambda \approx 400$ measured using nanoscale (black) and conventional (gray) probes. 
Exponents obtained from the local slope (local, squares) and with Extended Self-Similarity (ESS, circles) are compared with the She–Leveque (SL) model \cite{she1994universal}, the Sreenivasan–Yakhot (SY) model \cite{sreenivasan2021dynamics}, and the Kolmogorov 1941 prediction $\zeta_n = n/3$ (dotted line). 
Error bars denote standard errors over the inertial range. 
At high $Re_\lambda$, the exponents are consistent with saturation at high orders, 
approaching $\zeta_\infty \approx 2.2 \pm 0.1$ (horizontal gray line), whereas no such tendency is observed at $Re_\lambda \approx 400$. 
(b) Compensated structure functions, $\langle\Delta u^n\rangle r^{-\zeta_\infty}$ with $\zeta_\infty = 2.2$ at $Re_\lambda \approx 1400$. 
For $n \gtrsim 12$, the plateaus in the inertial range are parallel as expected. 
Colors as in Fig.~\ref{fig:Figure1}.} 
   \label{fig:Figure2}
\end{figure}

Turbulent structures can be statistically quantified with longitudinal structure functions, 
$S_n(r)$, which characterize the scale dependence of velocity increments. 
Figure~\ref{fig:Figure1} shows $S_n(r)$ for $2 \leq n \leq 14$. 
Over an intermediate range of separations, the data exhibit approximate power-law scaling, $S_n(r) \sim r^{\zeta_n}$, consistent with inertial-range dynamics. 
The higher-order exponents deviate systematically from $\zeta_n = n/3$, reflecting intermittency \cite{frisch1995turbulence}. 
We identified the inertial range with Kolmogorov’s four-fifths law, $\langle (\Delta u)^3 \rangle = -\frac{4}{5}\epsilon r$, by locating the region where the local slope $d\log |\langle (\Delta u)^3 \rangle|/d\log r \approx 1$, yielding an inertial range in the interval $0.02 \lesssim r/L \lesssim 0.5$ (see Fig.~\ref{fig:Figure1} (\textit{inset})), which extends to relatively large scales here, possibly due to dissipative effects \cite{sinhuber2017dissipative} or imposed shear \cite{ruiz2000scaling}. We extracted scaling exponents with local slopes, $\zeta_n(r) = d\log S_n/d\log r$, 
and with Extended Self-Similarity (ESS) \cite{benzi1993extended}. 
In the later case, we used the second-order structure function as a reference and set $\zeta_2=0.67$. 
The two methods yield exponents that agree up to $n=14$ (Fig.~\ref{fig:Figure2}\textit{a}). 

The measured exponents (Fig.~\ref{fig:Figure2}a) exhibit the expected concave dependence on moment order $n$, but depart systematically from phenomenological predictions~\cite{she1994universal,sreenivasan2021dynamics} beyond $n \approx 6$. 
At $Re_\lambda \approx 1400$ and for $n \gtrsim 12$, the exponents level off at a constant value, $\zeta_n \approx 2.2 \pm 0.1$, consistent with saturation of the exponents. The quoted uncertainty corresponds to the square root of the mean inertial-range variance of $\zeta_n$, averaged over $10 \leq n \leq 14$ and over both local-slope and extended self-similarity (ESS) estimates. This behavior is absent at the lower Reynolds number ($Re_\lambda \approx 400$), suggesting that saturation emerges only at sufficiently high $Re_\lambda$. 
At the lower $Re_\lambda$, measurements obtained with nanoscale and conventional probes are in agreement, indicating negligible spatial filtering by the probes over the scales resolved in common. 
At the larger $Re_\lambda$, however, the conventional probe lacks the spatial resolution required to capture the finer scales and is not shown here. 
Finally, for large $n$, the measured exponents lie systematically below phenomenological predictions, consistent with prior observations in shear-dominated turbulence \cite{casciola2005scaling,toschi1999intermittency,ruiz2000scaling,attili2012statistics}.

%\section{Evidence of Saturation of scaling exponents}

To evaluate the saturation of $\zeta_n$, we examine two quantities: 
(i) the compensated structure functions 
and (ii) the scale-dependent probability density functions (PDFs) of velocity increments. 
First, we consider the structure functions themselves. 
If the scaling exponent saturates at a finite value $\zeta_\infty$, then for sufficiently large order the compensated quantity $S_n(r) r^{-\zeta_\infty}$ should become independent of scale, $r$, within the inertial range. 
Figure \ref{fig:Figure2}\textit{b} shows the compensated structure functions using $\zeta_\infty =2.2$. 
For orders $n \gtrsim 12$, the curves display a plateau over the inertial range, consistent with the scaling exponent approaching a constant value for $n>12$.

An alternative test of saturation of the scaling exponents lies in the PDFs of velocity increments themselves, rather than in their moments. 
If the scaling exponents saturate at $\zeta_\infty$, 
then the tails of compensated PDFs, $Y(\Delta u (r)) \equiv r^{-\zeta_\infty}P(\Delta u(r))$, 
should collapse for separations within the inertial range \cite{celani2000universality}. 
High-order moments of the increments are dominated by the tails of the distributions, and are therefore sensitive to rare and extreme events. 
The scale-independence of the tails of the PDFs is less sensitive to statistical convergence than the moments of the PDFs \cite{celani2000universality,staicu2003small}. 

\begin{figure} %[H]
 \centering
 \includegraphics[width=0.9\linewidth, trim={10 155 12 105},clip]{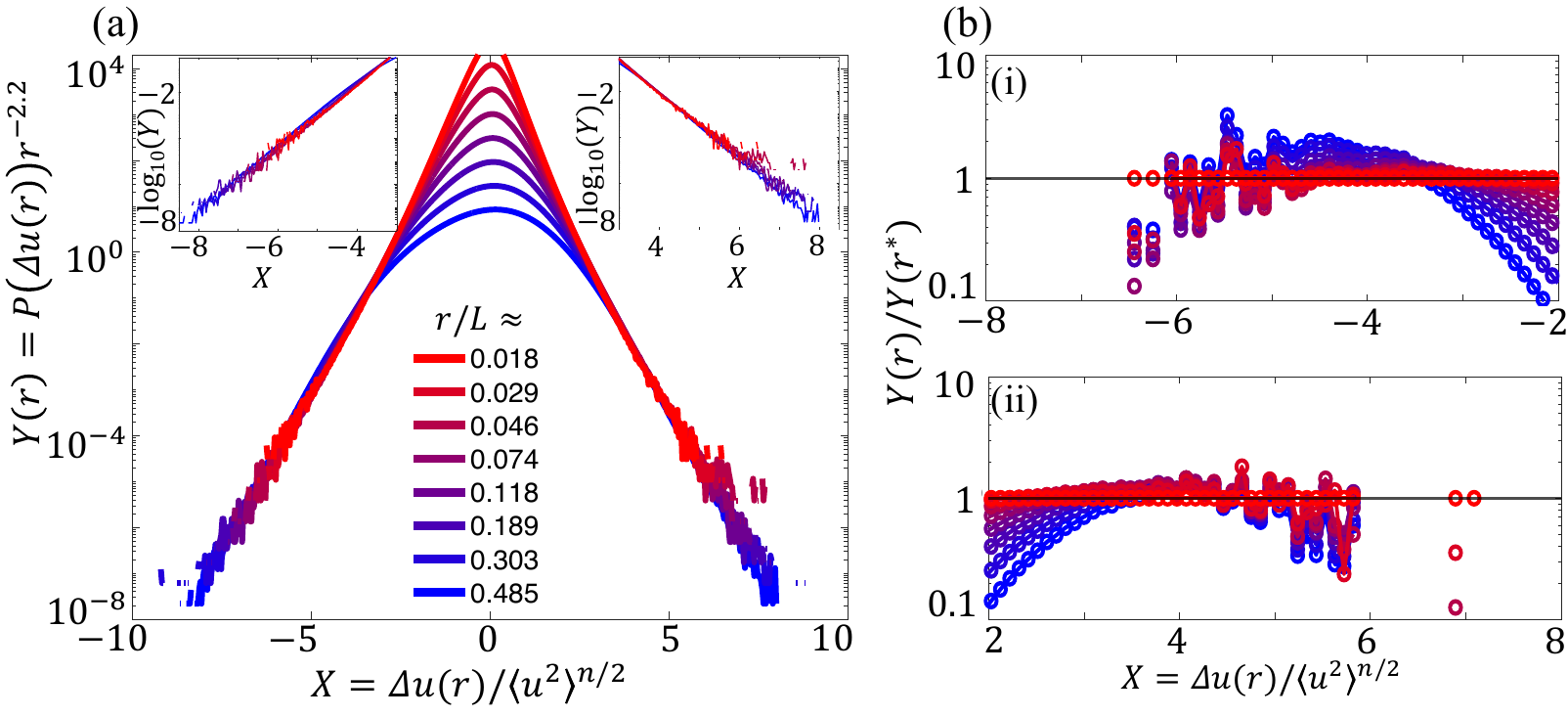}
    \caption{(a) Compensated PDFs, $Y(r)= P(\Delta u(r)) r^{-\zeta_\infty}$, of longitudinal velocity increments, evaluated using $\zeta_\infty=2.2$. 
The tails of the compensated PDFs collapse for $|\Delta u (r)|/\langle u^2\rangle^{1/2} \gtrsim 3.5$, nearly independent of the scale, $r$. 
The inset shows a magnified view of the far tails highlighting this collapse. 
(b) Ratios of the compensated PDF tails relative to the reference separation $r^* \equiv r/L\approx 0.02$. 
The ratios remain close to unity in the tail region, indicating collapse of the distributions across scales. 
This behavior supports the interpretation that the high-order structure-function exponents approach a constant value, $\zeta_p \to \zeta_\infty=2.2$, consistent with saturation of the scaling exponents at large order. } 
   \label{fig:Figure3}
\end{figure}

Figure \ref{fig:Figure3}\textit{a} shows the compensated PDFs for separations in the inertial range. 
The curves collapse for large normalized increments, indicating that the statistics in the tails become effectively scale-independent once compensated by $r^{-\zeta_\infty}$. 
To quantify this collapse, we plot the ratio $Y(\Delta u (r))/Y(\Delta u (r^*))$, where we chose the reference separation to be $r^* = 0.02\,L$. 
This ratio remains close to unity for both positive and negative tails when $|\Delta u (r)| \gtrsim 3.5 \langle u^2\rangle^{1/2}$ (see Fig.~\ref{fig:Figure3}\textit{b}). 
The deviations are typically within a factor of two. 
At even larger increment amplitudes the ratio begins to scatter, but this regime lies well beyond the increment amplitudes that contribute significantly to the moments up to order $n=14$ (discussed in the following section). 
Taken together, the flattening of the exponent curve, the plateau of the compensated structure functions, and the collapse of the compensated PDF tails all provide consistent evidence that the longitudinal structure-function exponents approach a finite saturation value, $\zeta_n \to \zeta_\infty \approx 2.2$. 

%\section{statistical convergence of moments}

\begin{figure} %[H]
 \centering
 \includegraphics[width=0.9\linewidth, trim={10 50 15 100},clip]{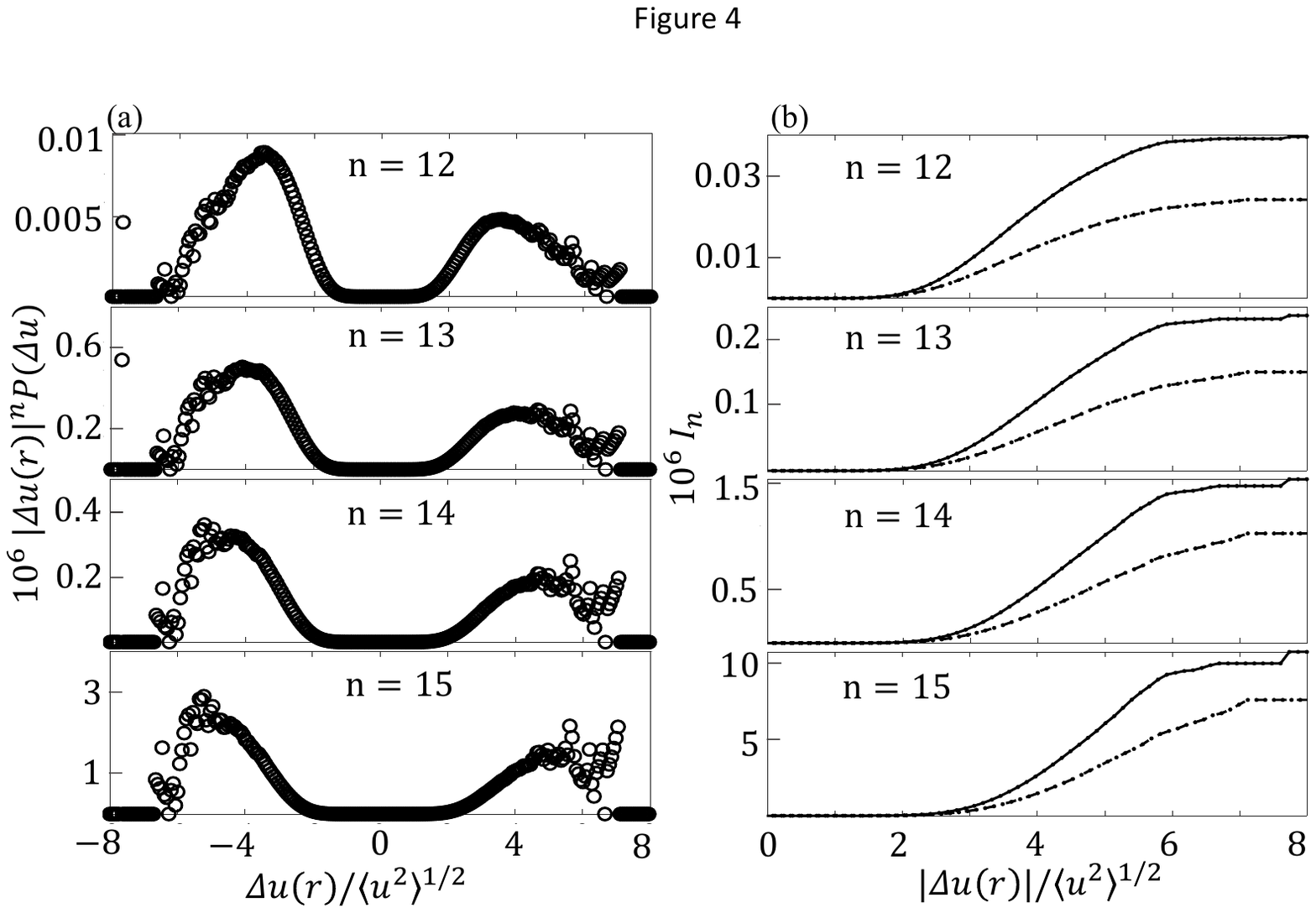}
    \caption{(a) Integrand of the $n$th-order moments of longitudinal velocity increments, 
$\Delta u^n P(\Delta u)$, for $12 \le n \le 15$, evaluated at the smallest inertial-range separation $r \approx 0.02L$. 
(b) Corresponding cumulative integrals of the moment contributions from the positive (dashed lines) and negative (solid lines) increment tails. For positive tails, $I_n=\int_{0}^{\infty} \Delta u^n P(\Delta u)\,du$  and for the negative tails, $I_n=\int_{-\infty}^{0} |\Delta u|^n P(\Delta u)\,du$. 
For $n \lesssim 14$, the cumulative integrals approach constant values at large increments, indicating that the dominant contributions from the far tails are statistically converged. } 
   \label{fig:Figure4}
\end{figure}

The apparent saturation of high-order exponents must be distinguished from artifacts due to undersampling of rare events, which can bias $\zeta_n$ at large $n$ \cite{frisch1995turbulence}. 
We assess convergence using the integrand $(\Delta u)^n P(\Delta u)$ (up to $n=15$) at $r \approx 0.02L$ (Fig.~\ref{fig:Figure4}\textit{a}). 
For $n \leq 14$, the integrands are peaked and decay rapidly, indicating that the dominant contributions arise from well-resolved increments. 
Deviations appear only for $n \gtrsim 14$ at extreme increments beyond the peaks, reflecting isolated rare events, but these contribute negligibly to moments up to $n=14$~\cite{staicu2003small}. 
Consistently, cumulative tail contributions (Fig.~\ref{fig:Figure4}\textit{b}) approach constant values for $n \leq 14$, with fluctuations confined to the extreme far tails. 
We therefore conclude that the structure functions are statistically converged up to $n=14$, and that the observed saturation of $\zeta_n$ is intrinsic rather than a sampling artifact.

%%%%%%%%%
 % \section{Discussion and conclusion}

The results provide evidence that at higher Reynolds numbers than previously observed in shear layers ($Re_\lambda \approx 1400$), the scaling exponents of longitudinal velocity structure functions approach a finite asymptotic value, $\zeta_n \to \zeta_\infty \approx 2.2$ for $n \gtrsim 12$. 
While anomalous scaling is a well-established signature of intermittency, saturation represents its extreme limit, indicating that high-order statistics are dominated by the most intense and spatially localized velocity fluctuations. 
Such behavior is known in systems where singular structures control the dynamics, 
including Burgers turbulence~\cite{bec2007burgers}, 
and passive scalar turbulence~\cite{celani2000universality,moisy2001passive,iyer2018steep}. 
In three-dimensional turbulence, prior evidence has been largely restricted to transverse increments~\cite{staicu2003small,iyer2018steep,buaria2023saturation}. 
The present results suggest that longitudinal increments also exhibit saturation at sufficiently high order, providing, to our knowledge, the first experimental evidence of this behavior in turbulence.

At a lower Reynolds numbers, $Re_\lambda \approx 400$, the scaling exponents did not exhibit a tendency toward saturation over the range of moments accessible in the present dataset. 
Instead, the exponents continued to increase with moment order without the flattening observed at $Re_\lambda \approx 1400$, suggesting that exponent saturation emerges only at Reynolds numbers large enough that intermittency is sufficiently pronounced, and possibly large enough for the statistics to be universal with respect to $Re_\lambda$ \cite{kuchler2023universal}.

The observed value for the largest scaling exponent, $\zeta_\infty \approx 2.2$, constrains the geometry of the most singular structures. 
Within the multifractal framework \cite{frisch1995turbulence}, saturation corresponds to a minimum H\"older exponent that is zero, yielding $\zeta_\infty = 3 - D(0)$. 
The present measurements therefore imply $D(0) \approx 0.8$, consistent with nearly one-dimensional structures. 
This is in line with the canonical picture of intense vortex filaments in turbulent flows \cite{douady1991direct,she1990intermittent,jimenez1993structure}, suggesting that such structures dominate the extreme statistics of longitudinal velocity increments. 
Notably, the She–Leveque model \cite{she1994universal}, while based on filamentary structures, predicts exponents that grow indefinitely due to $h_{\min}=1/9>0$; in contrast, the present results are consistent with $h_{\min}=0$. 
%, implying the dynamics in a distinct intermittency regime characterized by exponent saturation. 

If extreme events are indeed governed by filamentary structures, their characteristic scale must approach the Kolmogorov length~\cite{celani2000universality}. 
While the present nanoscale measurements resolve velocity increments down to dissipative scales, direct identification of the underlying structures requires multi-point or multi-component measurements capable of resolving the full velocity-gradient tensor. 
Nevertheless, the convergence of multiple independent diagnostics presented here provides self-consistent evidence that longitudinal structure-function exponents saturate at high order, thereby constraining the geometry of the most singular structures governing the turbulent cascade.

%\begin{acknowledgments}
We gratefully acknowledge Edmund T. Liu for his assistance with nano hot-wire manufacturing. This work was made possible through Office of Naval Research Grant (N00014-22-1-2038).
%\end{acknowledgments}

\bibliography{apssamp}% Produces the bibliography via BibTeX.

\end{document}